\documentclass[11pt]{article} 
\usepackage{amsmath}
\usepackage{amsfonts,amssymb}

\begin{document}

\newcommand{\nl}{\nonumber\\}
\newcommand{\nnl}{\nl[6mm]}
\newcommand{\nle}{\nl[-2.5mm]\\[-2.5mm]}
\newcommand{\nlb}[1]{\nl[-2.0mm]\label{#1}\\[-2.0mm]}
\newcommand{\ab}{\allowbreak}

\newcommand{\e}{{\mathrm e}}
\newcommand{\mm}{{\mathbf m}}
\newcommand{\nn}{{\mathbf n}}

\newcommand{\be}{\bes}
\newcommand{\ee}{\ees}
\newcommand{\bes}{\begin{eqnarray}}
\newcommand{\ees}{\end{eqnarray}}
\newcommand{\eens}{\nonumber\end{eqnarray}}

\renewcommand{\d}{\partial}

\newcommand{\eps}{\epsilon}
\newcommand{\si}{\sigma}
\newcommand{\ka}{\kappa}
\newcommand{\la}{\lambda}
\newcommand{\om}{\omega}

\newcommand{\vect}{{\mathfrak{vect}}}
\newcommand{\map}{{\mathfrak{map}}}
\newcommand{\gl}{{\mathfrak{gl}}}
\newcommand{\ssl}{{\mathfrak{sl}}}

\newcommand{\tr}{{\rm tr}}
\newcommand{\oj}{{\mathfrak g}}
                                            
\newcommand{\TT}{{\mathbb T}}
\newcommand{\RR}{{\mathbb R}}
\newcommand{\CC}{{\mathbb C}}
\newcommand{\ZZ}{{\mathbb Z}}
\newcommand{\NN}{{\mathbb N}}

\title{{Covariant Mickelsson-Faddeev extensions of gauge and diffeomorphism algebras}}

\author{T. A. Larsson \\
Vanadisv\"agen 29, S-113 23 Stockholm, Sweden\\
email: thomas.larsson@hdd.se}

\maketitle 
\begin{abstract} 
We construct new extensions of current and diffeomorphism algebras in $N>3$
dimensions, which are related to the Mickelsson-Faddeev algebra. The result is
compatible with Dzhumadil'daev's classification of diffeomorphism cocycles. We also
construct an extension of the current algebra in $N\geq5$ dimensions which
depends on the fourth Casimir operator.
\end{abstract}

\vskip 1cm

\section{Introduction}
Higher-dimensional analogs of affine and Virasoro algebras have been known for
a long time \cite{Kas85,Lar91,MRY90, PS88,RM94}. It is natural to ask whether
the algebras that pertain to gauge and diff anomalies in three dimensions, in particular
the Mickelsson-Faddeev (MF) algebra \cite{Fa84, Mi85, Mi89}, can also be generalized
to higher dimensions. The answer turns out to be affirmative, and is described
in the present paper in a Fourier basis on the $N$-dimensional torus. 
The key step is to replace the three-dimensional delta function with an
operator, which can be interpreted as an integral over a $3$-torus embedded in an 
$N$-torus. The resulting extensions are covariant in the sense that there is an 
intertwining action of $N$-dimensional diffeomorphisms. 

Given a covariant extension of the current algebra for $\gl(N)$, there is a standard
procedure to construct an associated extension of the diffeomorphism algebra
in $N$ dimensions. We apply this construction to the covariant MF extension, and
relate the result to Dzhumadil'daev's classification of cocycles for the algebra
of vector fields in $N$ dimensions \cite{Dzhu96}.

It is known that new gauge anomalies arise in all odd dimensions, which gives
rise to a hierarchy of extensions of the current algebra. Well-defined extensions 
involving the $n$:th Casimir operator are expected to arise in all dimensions 
$N \geq 2n-3$. We explicitly describe
the next element in this hierarchy, a covariant fourth-Casimir extension in $N \geq 5$
dimensions. This is presumably the simplest example of a non-abelian
extension. There is some overlap between \cite{CFNW94} and this part of
the present work, but the covariantization which leads to well-defined extensions 
above the minimal dimension $N=2n-3$ is new.

The final section contains a discussion on representations and possible relevance to physics.

\section{Multi-dimensional affine algebra}
\label{sec:affine}
Let $\oj$ be a finite-dimensional Lie algebra with generators $J^a$
and totally antisymmetric structure constants $f^{abc}$. The brackets 
are given by $[J^a,J^b] = if^{abc}J^c$.
Denote the symmetric Killing metric (proportional to the quadratic 
Casimir operator) by $\delta^{ab} = \tr\ J^aJ^b$. It is not necessary
to distinguish between upper and lower $\oj$ indices due to this 
metric; in contrast, the distinction is important for spacetime indices.
The current algebra $\map(M,\oj)$ is the algebra of maps from a
manifold $M$ to $\oj$. We specialize to the $N$-torus $\TT^N$ and 
expand all fields in a Fourier basis. The generators of
$\map(\TT^N,\oj)$ are $J^a(m) = \exp(im\cdot x) J^a$, where 
$m \equiv (m_0, m_1, ..., m_{N-1}) \in \ZZ^N$.

As is well known, $\map(\TT^1,\oj)$ admits a central extension,
the affine Kac-Moody algebra $\widehat\oj$:
\be
[J^a(m), J^b(n)] &=& if^{abc} J^c(m+n) + km \delta^{ab} \delta(m+n),
\label{KM}
\ee
where $\delta(m)$ is the Kronecker delta.
This extension is immediately generalized to $N>1$ \cite{Kas85,PS88}:
\be
[J^a(m), J^b(n)] &=& if^{abc} J^c(m+n) + k^\mu m_\mu \delta^{ab} \delta(m+n), 
\label{central}
\ee
where $k^\mu$ is a constant vector, e.g. $k^\mu = \delta^\mu_0$.
This formulation is not covariant because there is a privileged 
direction $k^\mu$. To remedy this defect, note that the Kronecker delta
in one dimension can be written as 
$\delta(m) = \frac1{2\pi} \int \exp(imx) dx$. The natural generalization
to $N$ dimensions is to replace the Kronecker delta with the curve operators
\be
S^\mu(m) = \frac1{2\pi} \int \e^{im\cdot x} dx^\mu,
\ee
where the integral is taken over some curve embedded in $\TT^N$.
These operators satisfy the constraint
\be
m_\mu S^\mu(m) = \frac1{2\pi i} \int d(\e^{im\cdot x}) \equiv 0.
\ee
The explicit form of $S^\mu(m)$ will not be important in the sequel.

The covariant form of the central extension (\ref{central}) is
defined by the relations
\bes
[J^a(m), J^b(n)] &=& if^{abc} J^c(m+n)
  + k\delta^{ab} m_\rho S^\rho(m+n), \nl  
{[}J^a(m), S^\mu(n)] &=& [S^\mu(m), S^\nu(n)] = 0, 
\label{KMN}\\
m_\mu S^\mu(m) &\equiv& 0.
\eens
This algebra is covariant in the sense that it admits an intertwining
action of general diffeomorphisms. Let $\vect(\TT^N)$ be the
algebra of vector fields on $\TT^N$, i.e. the algebra of 
infinitesimal diffeomorphisms. The generators 
$L_\mu(m) = -i \exp(i m\cdot x) \d/{\d x_\mu}$ satisfy
\be
[L_\mu(m), L_\nu(n)] &=& n_\mu L_\nu(m+n) - m_\nu L_\mu(m+n).
\label{vect}
\ee
The following brackets complete the definition of the semi-direct product \break
$\vect(\TT^N) \ltimes \map(\TT^N,\oj)$:
\bes
[L_\mu(m), J^a(n)] &=& n_\mu J^a(m+n), \nle
{[}L_\mu(m), S^\nu(n)] &=& n_\mu S^\nu(m+n)
 + k \delta^\nu_\mu m_\rho S^\rho(m+n).
\eens
$J^a(m)$ transforms as a scalar density of weight $+1$ and $S^\mu(m)$
as a vector density.

To prove that the extension (\ref{KMN}) is non-trivial it suffices to prove
non-triviality for some subalgebra. Fix a constant vector $e = (e_\mu)$. The generators 
$J^a(t) = J^a(te)$ satisfy the one-dimensional affine algebra (\ref{KM}) with
central charge $k = k^\mu e_\mu$. Hence the restriction of the extension to
this subalgebra is non-trivial.

\section{Multi-dimensional Mickelsson-Faddeev algebra}
\label{sec:MF}
The current algebra in three dimensions, $\map(\TT^3,\oj)$, admits a 
different type of extension, the Mickelsson-Faddeev (MF) algebra
\cite{Fa84,Mi85,Mi89}:
\bes
[J^a(\mm), J^b(\nn)] &=& if^{abc} J^c(\mm+\nn)
  + d^{abc} \epsilon^{ijk} m_i n_j A^c_k(\mm+\nn), \nl  
{[}J^a(\mm), A^b_j(\nn)] &=& if^{abc} A^c_j(\mm+\nn) 
  + \delta^{ab} m_j \delta(\mm+\nn), 
\label{MF}\\
{[}A^a_i(\mm), A^b_j(\nn)] &=& 0,
\eens
where Latin indices and boldface denote three-dimensional vectors:
$\mm = (m_1,m_2,m_3) \in \ZZ^3$.  The totally anti-symmetric
constant $\eps^{ijk}$ may be viewed as a 
tensor density of weight $+1$. Further, $A^a_\mu(\mm)$ are the 
Fourier components of the gauge connection, and 
$d^{abc} = \tr\ \{J^a, J^b\}J^c$ are the totally symmetric structure
constants proportional to the third Casimir. Constancy of 
$d^{bcd}$ implies the condition:
\be
f^{abe} d^{ecd} + f^{ace} d^{bed} + f^{ade} d^{bce} = 0.
\label{3-Casimir}
\ee

In analogy with the previous section, we observe that the 
three-dimen\-sional Kronecker delta can be written as an integral over $\TT^3$.
In higher dimensions, we replace it by a $3$-volume operator, which is an integral 
over some $3$-torus embedded in $\TT^N$:
\be
S^{\mu\nu\rho}(m) = \frac1 V \int \e^{im\cdot x} dx^\mu dx^\nu dx^\rho,
\ee
where $V$ is the $3$-volume of the embedded torus.
$S^{\mu\nu\rho}(m)$ does not commute with diffeomorphisms when $N>3$,
but it does possess the crucial property $m_\rho S^{\mu\nu\rho}(m) \equiv 0$.
Moreover, we make the change of basis:
\be
d^{abc} \epsilon^{ijk} A^c_k(\mm) \Rightarrow A^{ab,ij}(\mm).
\ee
The multi-dimensional MF algebra is defined by the equations:
\bes
[J^a(m), J^b(n)] &=& if^{abc} J^c(m+n) 
 + m_\mu n_\nu A^{ab,\mu\nu}(m+n), \nl
{[}J^a(m), A^{bc,\mu\nu}(n)] &=& if^{abd} A^{dc,\mu\nu}(m+n)
 + if^{acd} A^{bd,\mu\nu}(m+n) \nl
&+& d^{abc} m_\rho S^{\mu\nu\rho}(m+n), \nl
m_\rho S^{\mu\nu\rho}(m) &=& 0, 
\label{MFN}\\
A^{ab,\mu\nu}(m) &=& A^{ba,\mu\nu}(m) = -A^{ab,\nu\mu}(m), \nl
S^{\mu\nu\rho}(m) &=&  -S^{\nu\mu\rho}(m) =  S^{\rho\mu\nu}(m).
\label{SSym}
\ees
All other brackets vanish: $[J,S] = [A,A] = [A,S] = [S,S] = 0$.

It is readily verified, using the relations (\ref{3-Casimir}), that 
(\ref{MFN}) is a well-defined Lie algebra in any number of dimensions 
$N \geq 3$. Moreover, there is an intertwining action of $\vect(\TT^N)$,
under which $J^a(m)$, $A^{ab,\mu\nu}(m)$ and  $S^{\mu\nu\rho}(m)$ transform
as densities of weight $+1$ and the appropriate tensor types.
Explicity, 
\bes
[L_\mu(m), J^a(n)] &=& n_\mu J^a(m+n), \nl
{[}L_\mu(m), A^{ab,\nu\rho}(n)] &=& n_\mu A^{ab,\nu\rho}(m+n)
 + \delta^\nu_\mu m_\si A^{ab,\si\rho}(m+n)\nl
&+& \delta^\rho_\mu m_\si A^{ab,\nu\si}(m+n), \\
{[}L_\mu(m), S^{\nu\rho\si}(n)] &=& n_\mu S^{\nu\rho\si}(m+n) 
 + \delta^\nu_\mu m_\tau S^{\tau\rho\si}(m+n) \nl
&+& \delta^\rho_\mu m_\tau S^{\nu\tau\si}(m+n)
 + \delta^\si_\mu m_\tau S^{\nu\rho\tau}(m+n), \nl
\eens
To my knowledge, the multi-dimensional MF algebra (\ref{MFN}) is new.

Non-triviality of the multi-dimensional MF extension is proved as for the
multi-dimensional affine algebra. Introduce a constant vector $e = (e_\mu)$.
The restriction of (\ref{MFN}) to the subalgebra generated by
$J^a(me)$ is the ordinary MF algebra (\ref{MF}), which
is a non-trivial extension of $\map(\TT^3, \oj)$.

The original MF algebra (\ref{MF}) is recovered as follows. In three
dimensions, the closedness condition $m_\rho S^{\mu\nu\rho}(m)  = 0$
has the unique solution
\be
S^{\mu\nu\rho}(m) = \eps^{\mu\nu\rho}\delta(m),
\ee
and 
\be
A^a_\mu(m) = \eps_{\mu\nu\rho} d^{abc} A^{bc,\nu\rho}(m)
\ee
transforms as a connection.

\section{Multi-dimensional Virasoro algebra}
\label{sec:Vir}
Consider the current algebra $\map(\TT^N,\gl(N))$, with brackets
\be
[T^\mu_\nu(m), T^\rho_\si(n)] &=& \delta^\rho_\nu T^\mu_\si(m+n)
 - \delta^\mu_\si T^\rho_\nu(m+n).
\label{mapgl}
\ee
As usual, the current algebra generators transform as scalar densities
of weight $+1$:
\be
[L_\mu(m), T^\nu_\si(n)] &=& n_\mu T^\nu_\si(m+n).
\ee
There is an embedding $\vect(\TT^N) \hookrightarrow \vect(\TT^N) \ltimes \map(\TT^N,\gl(N))$,
defined by
\be
L'_\mu(m) = L_\mu(m) + m_\nu T^\nu_\mu(m).
\label{Lemb}
\ee
It is readily verified that the $L'_\mu(m)$ satisfy (\ref{vect}) if the unprimed 
$L_\mu(m)$ do so, and that the $T^\mu_\nu(m)$ transform as tensor densities:
\[
[L'_\mu(m), T^\nu_\rho(n)] = n_\mu T^\nu_\rho(m+n)
 + \delta^\nu_\mu m_\si T^\si_\rho(m+n) - m_\rho T^\nu_\mu(m+n).
\]
The embedding (\ref{Lemb}) is useful to create diffeomorphism analogs of current
algebra concepts. E.g., given the $\vect(N)$ module of scalar fields and a
$\gl(N)$ representation $R$, it produces the module of tensor densities of
type $R$. 

We will use the embedding to construct $\vect(\TT^N)$ extensions from \break
$\map(\TT^N,\gl(N))$ extensions. To construct the diffeomorphism analog
of the affine cocycle in section \ref{sec:affine}, we need the Killing metric
for $\gl(N)$. For a generic representation, the second Casimir must be of the form
\be
\delta^{\mu\rho}_{\nu\si} = \tr\ T^\mu_\nu T^\rho_\si
= c_1 \delta^\mu_\si \delta^\rho_\nu + c_2 \delta^\mu_\nu \delta^\rho_\si,
\ee
for some constants $c_1$ and $c_2$. Clearly, 
$\delta^{\rho\mu}_{\si\nu} = \delta^{\mu\rho}_{\nu\si}$.
In particular, the $\ssl(N)$ subalgebra of $\gl(N)$ is characterized by the
condition $T^\mu_\mu = 0$, which leads to the relation $c_1 + Nc_2 = 0$.

Now consider the multi-dimensional affine algebra (\ref{KMN}) in the 
particular case $\oj = \gl(N)$, and make use of the embedding (\ref{Lemb}).
The result is the multi-dimensional Virasoro algebra:
\bes
[L_\mu(m), L_\nu(n)] &=& n_\mu L_\nu(m+n) - m_\nu L_\mu(m+n) \nl 
&+& (c_1 m_\nu n_\mu + c_2 m_\mu n_\nu) m_\rho S^\rho(m+n), \nl
{[}L_\mu(m), S^\nu(n)] &=& n_\mu S^\nu(m+n)
 + \delta^\nu_\mu m_\rho S^\rho(m+n), 
\nlb{Vir}
{[}S^\mu(m), S^\nu(n)] &=& 0, \nl
m_\mu S^\mu(m) &=& 0.
\eens
To see that this algebra indeed reduces to the usual Virasoro algebra when
$N=1$, we notice that the condition $m_0 S^0(m_0)$ implies that 
$S^0(m_0)$ is proportional to the Kronecker delta, which indeed commutes
with diffeomorphisms. So the Virasoro extensions is central when $N=1$
but not otherwise. Nevertheless, (\ref{Vir}) defines a well-defined and
non-trivial Lie algebra extension of $\vect(N)$ for every $N$.

The cocycle proportional to $c_1$ was discovered by Rao and Moody 
\cite{RM94}, and the one proportional to $c_2$ by myself \cite{Lar91}.

\section{Multi-dimensional MF-diffeomorphism algebra}
\label{sec:MFvect}
In this section we intend to use the embedding (\ref{Lemb}) to construct
an analog of the MF extension for $\vect(\TT^N)$. To this end, we first
need to specialize the multi-dimensional MF algebra (\ref{MFN}) to
$\oj = \gl(N)$:
\bes
[T^\mu_\nu(m), T^\rho_\si(n)] &=& \delta^\rho_\nu T^\mu_\si(m+n)
 - \delta^\mu_\si T^\rho_\nu(m+n) \nl
 &+& m_\ka n_\la R^{\mu\rho,\ka\la}_{\nu\si}(m+n), 
\nlb{MFgl}
{[}T^\mu_\nu(m), R^{\rho\tau,\ka\la}_{\si\om}(n)] &=&
 \delta^\rho_\nu R^{\mu\tau,\ka\la}_{\si\om}(m+n)
 - \delta^\mu_\si R^{\rho\tau,\ka\la}_{\nu\om}(m+n) \nl
&+&\delta^\tau_\nu R^{\rho\mu,\ka\la}_{\si\om}(m+n)
 - \delta^\mu_\om R^{\rho\tau,\ka\la}_{\si\nu}(m+n) \nl
&+&\delta^\ka_\nu R^{\rho\tau,\mu\la}_{\si\om}(m+n)
 + \delta^\la_\nu R^{\rho\tau,\ka\mu}_{\si\om}(m+n) \nl
&+&d^{\mu\rho\tau}_{\nu\si\om} m_\pi S^{\ka\la\pi}(m+n), \nl
R^{\mu\rho,\ka\la}_{\nu\si}(m) &=&
  R^{\rho\mu,\ka\la}_{\si\nu}(m) =
  -R^{\mu\rho,\la\ka}_{\nu\si}(m),
\label{RSymm}
\ees
where $d^{\mu\rho\tau}_{\nu\si\om}$ is the third Casimir for $\gl(N)$.
For symmetry reasons it must be of the form
\bes
d^{\mu\rho\tau}_{\nu\si\om} &=& 
 a_1 (\delta^\mu_\si \delta^\rho_\om \delta^\tau_\nu
  + \delta^\mu_\om \delta^\rho_\nu \delta^\tau_\si) \nl
&+&a_2( \delta^\mu_\nu \delta^\rho_\om \delta^\tau_\si
+ \delta^\mu_\om \delta^\rho_\si \delta^\tau_\nu
+ \delta^\mu_\si \delta^\rho_\nu \delta^\tau_\om) 
\label{glNCasimir3} \\
&+&a_3 \delta^\mu_\nu \delta^\rho_\si \delta^\tau_\om.
\eens
In particular for $\ssl(N) \subset \gl(N)$, the condition
$d^{\mu\rho\tau}_{\nu\si\tau} = 0$ leads to $a_1 = N^2 a$, $a_2 = -2N a$,
$a_3 = 4a$. We check that the third Casimir vanishes for the trivial
algebra $\ssl(1)$.

The MF extension of $\vect(\TT^N)$ becomes:
\bes
[L_\mu(m), L_\nu(n)] &=& n_\mu L_\nu(m+n) - m_\nu L_\mu(m+n) \nl 
&+&m_\rho m_\ka n_\si n_\la R^{\rho\si,\ka\la}_{\mu\nu}(m+n), \nl
{[}L_\mu(m), R^{\rho\tau,\ka\la}_{\si\om}(n)] &=& 
 n_\mu R^{\rho\tau,\ka\la}_{\si\om}(m+n) \nl
&+&\delta^\rho_\mu m_\nu R^{\nu\tau,\ka\la}_{\si\om}(m+n)
 - m_\si R^{\rho\tau,\ka\la}_{\mu\om}(m+n) \nl
&+&\delta^\tau_\mu m_\nu R^{\rho\nu,\ka\la}_{\si\om}(m+n)
 - m_\om R^{\rho\tau,\ka\la}_{\si\mu}(m+n) \nl
&+&\delta^\ka_\mu m_\nu R^{\rho\tau,\nu\la}_{\si\om}(m+n)
 + \delta^\la_\mu m_\nu R^{\rho\tau,\ka\nu}_{\si\om}(m+n) \nl
&+&d^{\nu\rho\tau}_{\mu\si\om} m_\nu m_\pi S^{\ka\la\pi}(m+n),
\label{MFvect}\\
{[}L_\mu(m), S^{\nu\rho\si}(n)] &=& n_\mu S^{\nu\rho\si}(m+n)
+ \delta^\nu_\mu m_\tau S^{\tau\rho\si}(m+n) \nl
&+& \delta^\si_\mu m_\tau S^{\nu\tau\si}(m+n)
 + \delta^\rho_\mu m_\tau S^{\nu\rho\tau}(m+n), \nl
m_\rho S^{\mu\nu\rho}(m) &=& 0, 
\eens
All other brackets vanish; $[R,R] = [R,S] = [S,S] = 0$. The $R$ and $S$
symmetry properties were written down in (\ref{RSymm}) and (\ref{SSym}),
respectively. Using the expression (\ref{glNCasimir3}) for the $\gl(N)$ 
third Casimir, the final term in the $LR$ bracket can be written more 
explicitly as
\bes
&&\Big( a_1(m_\si\delta^\rho_\om\delta^\tau_\mu 
  + m_\om\delta^\rho_\mu\delta^\tau_\si)
+ a_2 (m_\mu\delta^\rho_\om\delta^\tau_\si
 + m_\om\delta^\rho_\si\delta^\tau_\mu
  + m_\si\delta^\rho_\mu\delta^\tau_\om) \nl
&&\qquad +\ a_3 m_\mu\delta^\rho_\si\delta^\tau_\om \Big)
 \times\ m_\nu S^{\ka\la\nu}(m+n).
\ees

Dzhumadil'daev has classified extensions of the algebra of polynomial 
vector fields by modules of tensor densities \cite{Dzhu96}. His 
classification applies morally to $\vect(M)$ in general, and
in particular to $\vect(\TT^N)$ which contains polynomial vector fields
as a proper subalgebra. The MF-diffeomorphism algebra (\ref{MFvect}) is 
closely related to cocycles $\psi^W_3 - \psi^W_{10}$ in his classification.
Namely, these cocycles are recovered if we set $S^{\mu\nu\rho}(m) \equiv 0$
(which can be consistently done because the bracket of $S$ with anything is 
proportional to $S$) and decompose $R^{\mu\rho,\ka\la}_{\nu\si}$ into
irreducible $\gl(N)$ representations. These cocycles are also described in
subsection 3.3 of the review \cite{Lar00}.

\section{A fourth Casimir extension of current algebras}

The descent equations suggest that there is an entire hierarchy of
extensions of $\map(\TT^N, \oj)$. In this section we explicitly describe
the next element in this hierarchy, associated with the fourth order Casimir operator.
Let
\be
d^{abcd} \equiv \tr (J^a J^b J^c J^d + \mathrm{permutations})
\ee
be the structure constants of the fourth Casimir operator 
of $\oj$. They satisfy the condition
\be
f^{abf} d^{fcde} + f^{acf} d^{bfde} + f^{adf} d^{bcfe} 
 + f^{aef} d^{bcdf} = 0,
\label{4-Casimir}
\ee
which follows from the fact that $d^{abcd}$ commutes with $\oj$.

The fourth Casimir extension of $\map(\TT^N, \oj)$ is defined by the
relations
\bes
[J^a(m), J^b(n)] &=& if^{abc} J^c(m+n) 
 + m_\mu n_\nu A^{ab,\mu\nu}(m+n), \nl
{[}J^a(m), A^{bc,\mu\nu}(n)] &=& if^{abd} A^{dc,\mu\nu}(m+n)
 + if^{acd} A^{bc,\mu\nu}(m+n) \nl
&+&m_\rho n_\si B^{abc,\mu\nu\rho\si}(m+n),
\nl
{[}J^a(m), B^{bcd,\mu\nu\rho\si}(n)] &=& 
 if^{abe} B^{ecd,\mu\nu\rho\si}(m+n)
 + if^{ace} B^{bed,\mu\nu\rho\si}(m+n) \nl
&+& if^{ade} B^{bce,\mu\nu\rho\si}(m+n)
 + kd^{abcd} m_\tau S^{\mu\nu\rho\si\tau}(m+n), 
\nl
{[}A^{ab,\mu\nu}(m), A^{cd,\rho\si}(n)] &=&
 i f^{ace} B^{bde,\mu\nu\rho\si}(m+n)
 + i f^{bce} B^{ade,\mu\nu\rho\si}(m+n) \nl
&+& i f^{ade} B^{bce,\mu\nu\rho\si}(m+n)
 + i f^{bde} B^{ace,\mu\nu\rho\si}(m+n) \nl
&+& kd^{abcd} m_\tau S^{\mu\nu\rho\si\tau}(m+n), \nl
m_\tau S^{\mu\nu\rho\si\tau}(m) &=& 0.
\label{MF4}
\ees
Note that the constants $k$ in the third and fourth equations are equal.
The symmetry properties can be summarized as
\bes
A^{ab,\mu\nu}(m) &=& A^{(ab),[\mu\nu]}(m), \nl
B^{adc,\mu\nu\rho\si}(m) &=& B^{(adc),[\mu\nu\rho\si]}(m),
\label{4sym}\\
S^{\mu\nu\rho\si\tau}(m) &=&  S^{[\mu\nu\rho\si\tau]}(m),
\eens
where $(...)$ denotes symmetrization and $[...]$ denotes anti-symmetrization
of indices. Finally, the structure constants satify the 
conditions (\ref{3-Casimir}) and (\ref{4-Casimir}).

To verify the Jacobi identities is straightforward albeit tedious. 

In the special case of five dimensions, the $5$-volume operator
is proportional to the Kronecker delta:
\be
S^{\mu\nu\rho\si\tau}(m) = \eps^{\mu\nu\rho\si\tau}\delta(m),
\ee
and 
\be
B^a_\mu(m) = \eps_{\mu\nu\rho\si\tau} d^{abc} B^{bc,\nu\rho\si\tau}(m)
\ee
transforms as a connection.

The algebra (\ref{MF4}) is an example of a non-abelian extension,
because the $[A,A]$ bracket is nonzero. There are some similarities to the
algebra described in equation (1.2) of \cite{CFNW94}.
However, their algebra is only defined in five dimensions, whereas
(\ref{MF4}) is well defined in all dimensions $N\geq5$, because the
five-dimensional Kronecker delta has been replaced by the
covariant $5$-volume operator $S^{\mu\nu\rho\si\tau}(m)$. Moreover,
their algebra only involves the third Casimir $d^{abc}$, so (\ref{MF4})
may be new even in $N=5$ dimensions.

It is clear from the existence of the embedding (\ref{Lemb}) that 
$\vect(N)$ possesses an analog extension for $N \geq 5$. The explicit
expressions will be quite cumbersome and not very illuminating, and
we have not written them down.

\section{Discussion}

In this paper we have indicated how the $n$:th Casimir extensions of the
current and diffeomorphism algebras in $N = 2n-3$ dimensions give
rise to extensions also in higher dimensions. The construction includes
the multi-dimensional affine algebras \cite{Kas85,PS88}, Virasoro 
algebras \cite{Lar91,RM94}, as well as the presumably new 
multi-dimensional MF and fourth Casimir algebras. The MF extension of
the current algebra in four dimensions could possibly be useful to
study gauge anomalies in a covariant formalism.

The price to pay is that we must introduce $(2n-3)$-volume operators to
replace the delta functions. In the minimal dimension, the embedded torus is
essentially unique (a circle can be embedded into the circle in one way
only), and the  sub-volume operator commutes with diffeomorphisms. Above
the minimal dimension $N = 2n-3$, this is no longer true. 

It is difficult to imagine what the physical meaning of such a sub-volume operator 
could be. Hence the algebras described in this paper do 
probably not occur in nature. This is further corroborated by that fact
that the MF algebra apparently lacks good quantum representations;
more precisely, it has no unitary lowest-weight representation acting 
on a separable Hilbert space \cite{Pic89}. Mickelsson has constructed
a different type of representations \cite{Mi92}, but since these involve a
classical background gauge field they can not arise in a fundamental, fully
quantum theory. Nature abhors algebras without unitary quantum representations, 
which is confirmed by the fact that gauge
anomalies associated with the MF algebra cancel in the standard model.

The situation is different for the multi-dimensional affine and Virasoro
algebras. In every real-world experiment, there is a naturally occurring one-dimensional curve -- the
observer's spacetime trajectory. Moreover, these algebras do possess unitary
lowest-weight representations. In fact, a classification of such 
representations for the multi-dimensional affine algebra appeared already
in chapter 4 of \cite{PS88}, and they have been further studied in e.g.
\cite{BB98,Bil97,Lar98,MRY90}. Instead of working with quantum fields,
one must consider spacetime histories in the space of $p$-jets, which locally
can be identified with Taylor series truncated at order $p$. The
privileged one-dimensional curve is identified as the time evolution of
the expansion point, i.e. the observer's trajectory in spacetime.
Because the space of $p$-jets is finite-dimensional, the space of
$p$-jet trajectories is spanned by finitely many functions of a single
variable. In this situation we can normal order operators without
producing infinities.

Since the multi-dimensional affine and Virasoro algebras do possess
unitary lowest-weight representations, nothing prevents them from appearing
in nature. Such an extension is a gauge or diff anomaly,
but that is by itself not a sign of inconsistency\footnote{Counterexample:
the free subcritical string, which according to the no-ghost theorem
can be quantized with a ghost-free spectrum despite its conformal 
gauge anomaly.}. A gauge anomaly 
simply means that the classical and quantum theories have different
symmetries. It is of course inconsistent to treat an anomalous gauge
symmetry as a redundancy; a gauge anomaly converts a classical gauge
symmetry into a quantum global symmetry, which acts on the Hilbert space
rather than reducing it. The crucial consistency condition is that the
symmetry algebra is represented unitarily.

However, these extensions can not arise within QFT. In four dimensions,
QFT gauge anomalies are proportional to the third Casimir $d^{abc}$,
and are hence described by the MF algebra rather than by the multi-dimensional
affine algebra. Moreover, there are no QFT diff anomalies at all in
four dimensions \cite{Bon86}. The reason is that the extensions of the
diffeomorphism algebra in four dimensions
involve the curve operator $S^\mu(m)$, which requires that the observer's
trajectory has been introduced. This suggests that in order to fully
understand gauge and general-covariant theories, QFT must be
amended with a physical observer with a quantized position operator.

\end{document}